# The north-south divide in the Italian higher education system[1]

Giovanni Abramo (corresponding author)
*Laboratory for Studies of Research and Technology Transfer*
*Institute for System Analysis and Computer Science (IASI-CNR)*
*National Research Council of Italy*
ADDRESS: Consiglio Nazionale delle Ricerche
Istituto di Analisi dei Sistemi e Informatica
Via dei Taurini 19, 00185 Roma - ITALY
tel. and fax +39 06 72597362, giovanni.abramo@uniroma2.it

Ciriaco Andrea D'Angelo
*Department of Engineering and Management-University of Rome "Tor Vergata"*
ADDRESS: Dipartimento di Ingegneria dell'Impresa
Università degli Studi di Roma "Tor Vergata",
Via del Politecnico 1, 00133 Roma - ITALY
tel. and fax +39 06 72597362, dangelo@dii.uniroma2.it

Francesco Rosati
*Department of Management Engineering-Technical University of Denmark*
ADDRESS: Technical University of Denmark
Centrifugevej, Building 372
2800 Kgs. Lyngby - Denmark
tel +45 45256021, frro@dtu.dk

## Abstract

This work examines whether the macroeconomic divide between northern and southern Italy is also present at the level of higher education. The analysis confirms that the research performance in the sciences of the professors in the south is on average less than that of the professors in the north, and that this gap does not show noticeable variations at the level of gender or academic rank. For the universities, the gap is still greater. The study analyzes some possible determinants of the gap, and provides some policy recommendations for its reduction.

## Keywords

---

## 1. Introduction

The problem of the north-south differential in Italian economic development has attracted the attention of politicians, economists, historians, geographers and scholars of cultural issues ever since the unification of the Italian state, in 1861 (Daniele and Malanima, 2011). Since then, the evident persistence of the north-south differential has fed heated political-economic and historical-geographical debate (e.g. Eckaus, 1961; Daniele and Malanima, 2007), centering mainly on the characteristics and causes of the differential, and obviously on the potential measures to be adopted to reduce its extent.

Other countries also experience gaps in regional development, however in Italy the dualism between north and south is particularly strong, and remains irresolvable despite the repeated interventions aimed at reducing the gap. Historians are of common opinion that the roots of the differential are remote in time, deriving from factors in the history, economy, political geography, culture and society of Italy and Europe (Daniele and Malanima, 2007). Niceforo (1901) ascribes the divide to anthropological factors. Banfield (1958), and Putnam et al. (1993) identify its roots in the lower public spirit and deployment of social capital in the south. Aprile (2010) hypothizes that the north has in fact exploited the south that, before the unification of Italy in 1870, was florid and developed. Franchetti and Sonnino (1877) hold that the unfavourable geographical position of the south plays a role. Daniele and Malanima (2014) argue that the economic divide is a consequence of the interplay of economic factors during the Italian industrialisation starting at the end of the 19th century. According to Felice (2014), instead, the problem lies with the ruling class that has allocated resources with the aim of maximizing economic rents rather than socio-economic development. Along the same line, Cannari et al. (2010) and Trigilia (2012) state that the south lags behind because of the local political class, that revealed unable or unwilling to promote social development.

Over time, this differential has undergone various mutations, becoming more pronounced in some phases and lessening in others (Eckaus, 1960, 1969; Esposto, 1992; 1997; Cafagna, 1989). A study by Daniele and Malanima (2007) shows that prior to Italian unification, the gaps between the different regions in terms of per capita production were insignificant, and that the first differences appeared with the onset of economic modernization, more or less between 1880 and the First World War, with the strong development of some of the north-western regions, situated closer to the industrializing European nations. The gap widened during the two decades of Fascist government, then lessened considerably between 1953 and 1973, as modern economic growth continued to expand, involving new territories and populations in both the north-center and south. Again according to Daniele and Malanima (2007), the gap then worsened with the slowing of economic growth rates from the early 1970s onwards. Finally, beginning in 2007, the recent economic crisis has amplified the gap to the point that the south now seems considered at risk of permanent underdevelopment (SVIMEZ, 2015). The latest regional economic statistics (ISTAT, 2015a) show that the per-capita GDP of 17,200 euro in the south is 46% less than that of the north-center. At 2013 prices, consumer spending was 18,300 euro per person in the north-center and 12,500 euro in the south. Concerning infrastructure, the latest Regional Competitiveness Index (RCI) – Infrastructure puts the southern Italian regions in the last places of the 259 NUTS 2 regions of the 28-nation EU (Annoni and Dijkstra, 2013).

According to Lynn (2010), the north-south differential is manifested not only in macroeconomic terms: the author in fact claims there are differences between north and south in terms of cognitive capacities, particularly IQ, and goes on to show how these cognitive differences predict further differences in income, education, infant mortality, height and literacy. Needless to say that Lynn's positions have ignited a heated debate which is still open. His study in fact has soon received strong criticism, in both conceptual and methodological terms (e.g. Felice and Giugliano, 2011; Cornoldi et al., 2010; Beraldo, 2010). The author then replied with new evidence (Lynn, 2012; Piffer & Lynn, 2014), backed also by Carl's (2014) study showing that the association between average trust and log GDP per capita across Italian regions disappears or attenuates substantially when regional IQ is controlled for. The latest work on the subject is by Daniele (2015), who objects to a possible nexus between IQ differences and socio-economic development.

Regardless of the nation or region, universities clearly play an important role in the support of industrial competitiveness, socio-economic development and social mobility. In the context of the north-south gap, the academic institutions can in fact contribute to reducing the differential, or vice versa to increasing it, both in macroeconomic and social terms. A balanced university offer in all the regions would then ensure equal opportunities for education and training, including for those less well-off, who cannot support the greater costs of an education "away from home". In addition, all the stakeholders situated in the territory could benefit from the knowledge spillovers from the universities' research activity. Indeed these effects are primarily local, because of the geographic proximity effect (Anselin et al., 1997; Jaffe, 1989; Autant-Bernard, 2001). However, for such spillovers to happen, it is necessary that the quality of professors be equally distributed through the national territory. If equal distribution is lacking, then the stakeholders of the education and research would whenever possible orient themselves towards the better universities. Until 2009, government policies were designed to guarantee equal access to higher education independent of the localization, including through the allocation of public resources based on the number of students and the type of teaching offer, rather than on the basis of merit. Among other effects, this inhibited any significant differentiation among Italian universities. In 2009 the situation began to change. At this time, following the first national evaluation exercise (VTR 2001-2003), the government adopted performance-based research funding (PBRF) of universities. This mechanism in itself would normally not be harmful, being founded on a logic of competitive assignment of resources, to stimulate continuous improvement. However, there could indeed be a problem if the better researchers were concentrated above all in the universities of the more developed geographic areas. In this case, the resources would move from the less developed to the more developed areas (in the Italian case, from south to north). This would have negative outcomes for the less developed areas, both in economic and social terms. According to some observers, PBRF could in fact cause a further increase in the north-south differential in macroeconomic and social development (Viesti, 2015, Abramo et al., 2011a). On the one hand, the PBRF policies can stimulate competitiveness and therefore the performance of the professors in the weaker universities. However, in decreasing the resources at the disposition of these same universities (often concentrated in the south), it could instead contribute to increasing the socio-economic differential, rather than diminishing it. What would arise then would be an education system that is ever more differentiated among institutions with more or less assets (in terms of finances, faculty, students, foreign contacts and relations), with the better-off universities strongly concentrated in certain areas of northern Italy.

In summary, in Italy there seem to be all the conditions for an accentuation of the north-south divide. In the current work, we intend to identify whether this divide is also present in the higher education system. To determine this, we will compare the scientific performance of the professors and of the universities in northern, central and southern Italy. We will carry out the analyses at the aggregate, discipline and field levels. We will examine all professors as of 31/12/2015, active in research in the Sciences, for whom bibliometrics can be applied to measure the research performance. We evaluate their scientific product indexed in the Web of Science (WoS) over the period 2009-2013. Finally, we will analyze the possible causes that may have contributed to the north-south differential in higher education, in particular the possible relations between the gap in higher education and the socio-economic gap. To the best of our knowledge, there are no other studies on the subject with the only exception of an istitutional report (which is not a study) by the Italian Agency for the Evaluation of University and Research (ANVUR, 2014). The report aggregates at territorial level the results of the national research assessment exercise VQR 2004-2010, which has been strongly criticized (Franco, 2013; Baccini and De Nicolao, 2016; Baccini, 2016; Abramo, D'Angelo, and Di Costa, 2014a; Abramo and D'Angelo, 2015a) in terms of methodology and performance indicators (one for all, the overall assessement of universities is based on three products only per professor over a seven year period).

The next section of the paper describes the structure of the Italian higher education system. Sections 3 and 4 present the methodology and dataset used for the analyses. In section 5 we present the results of the study. The final section offers the authors' discussion and conclusions.

## 2. The Italian higher education system

The Italian Ministry of Education, University and Research (MIUR) recognizes a total of 96 universities as having the authority to issue legally recognized degrees. Of these, 29 are small, private, special-focus universities, of which 13 offer only e-learning, 67 are public and generally multi-disciplinary universities, scattered throughout Italy. Six of them are *Scuole Superiori* (Schools for Advanced Studies), specifically devoted to highly talented students, with very small faculties and tightly limited enrolment per degree program. In the overall system, 94.9% of faculty are employed in public universities (0.5% in *Scuole Superiori*). Universities are largely financed by the government through non-competitive allocation of funds. Until 2009 the core government funding (56% of universities' total income) was input oriented (i.e. independent of merit, and distributed to universities in a manner intended to satisfy the needs of each and all equally, with respect to their size and research disciplines). It was only following the first national research evaluation exercise (VTR 2001-2003), conducted in the period 2004-6, that a minimal share, equivalent to 3.9% of total income, was assigned by the MIUR as a function of the assessment of research and teaching. Despite interventions intended to grant increased autonomy and responsibilities to the universities (Law 168 of 1989)[2], the Italian higher education

---

[2] This law was intended to grant increased autonomy and responsibility to the universities to establish their own organizational frameworks, including charters and regulations. Subsequently, Law 537 (Article

system is a long-standing, classic example of a public and highly centralized governance structure, with low levels of autonomy at the university level and a very strong role played by the central state.

In keeping with the Humboldtian model, there are no 'teaching-only' universities in Italy, as all professors are required to carry out both research and teaching. National legislation includes a provision that each faculty member must provide a minimum of 350 hours per year of teaching. At the close of 2015, there were 54,800 faculty members in Italy (full, associate and assistant professors) and a roughly equal number of technical-administrative staff. Salaries are regulated at the central level and are calculated according to role (administrative, technical or professorial), rank within role (e.g. assistant, associate or full professor) and seniority. None of a professor's salary depends on merit. Moreover, as in all Italian public administration, dismissal of unproductive employees is unheard of. All new personnel enter the university system through public competitions, and career advancement depends on further public competitions. In absence of a significant differentiation and salary leverage to attract talented professors, in theory there is no reason to expect an uneven geographical distribution of research performance of academic staff. However, the entire legislative-administrative context has created a culture that is hardly competitive, yet flourishing with favoritism and other opportunistic behaviors (Zagaria, 2007; Perotti, 2008).

Abramo et al. (2014b) investigated 287 associate professor competitions. The analysis showed several critical issues, particularly concerning unsuccessful candidates who outperformed the competition winners in terms of research productivity, as well as a number of competition winners who resulted as totally unproductive. Almost half of individual competitions selected candidates who would go on to achieve below-median productivity in their field of research over the subsequent triennium. A more recent work (Abramo et al., 2015a) showed that the fundamental determinant of an academic candidate's success is not scientific merit, but rather the number of years that the candidate has belonged to the same university as the president of the selection committee. Given the environment, Italian universities are unable to attract significant numbers of talented foreign faculty: only 1.8% of research staff are foreign nationals. Over the period 2009-2013, 3,178 (9.1%) of the 34,862 professors in the Sciences did not publish any scientific articles in WoS indexed journals. Another 868 professors (2.5%) achieved publication, but their work was never cited. This means that 4,046 individuals (11.6%) had no impact on scientific progress measurable by bibliometric databases. This share of unproductive faculty has been declining but is still too high, particularly given that the legislative structure obligates all professors to conduct research. Differently from competitive higher education systems, top scientists are not concentrated in a limited number of universities. Instead, they are dispersed more or less uniformly among all Italian universities, along with the unproductive academics (Abramo et al., 2012a), so that no single institution reaches the critical mass of excellence necessary to develop as an elite university and to compete internationally.

---

5) of 1993 and Decree 168 of 1996 provided further changes intended to increase university involvement in overall decision-making on use of resources, and to encourage individual institutions to operate in the market and reach their own economic and financial equilibrium.

## 3. The dataset

The dataset used for the analysis was extracted from http://cercauniversita.cineca.it, a service database gathered by the MIUR. The database includes assistant, associate and full professors of all Italian universities. In the Italian academic system, each professor is classified in one and only one research field. There are a total of 370 such fields (named scientific disciplinary sectors, or SDSs[3]), grouped into 14 disciplines (named university disciplinary areas), or UDAs.

For each record, the MIUR database shows: last name and first name, university, SDS, academic rank, department. The database is updated at the end of each year. At 31/12/2015 the database contained 54,800 records.

Our analysis is limited to professors working in 9 Sciences UDAs, where scientific performance can be assessed with an acceptable level of reliability using bibliometric techniques: Mathematics and computer science; Physics; Chemistry; Earth sciences; Biology; Medicine; Agricultural and veterinary sciences; Civil engineering; Industrial and information engineering.

The dataset therefore includes 34,856 professors (14,912 belonging to northern universities, 10,993 to southern ones). These professors belong to 90 different universities (34 in the north, 29 in the center and 27 in the south Italy).

Table 1 presents the distribution of the professors and universities by UDA per geographic macro-region.

| | Professors | | | | Universities | | | |
|---|---|---|---|---|---|---|---|---|
| UDA | North | Centre | South | Total | North | Centre | South | Total |
| Math & Computer | 1,522 (46.6) | 807 (24.7) | 939 (28.7) | 3,268 | 28 (37.8) | 25 (33.8) | 21 (28.4) | 74 |
| Physics | 1,015 (43.5) | 596 (25.5) | 722 (30.9) | 2,333 | 26 (39.4) | 17 (25.8) | 23 (34.8) | 66 |
| Chemistry | 1,336 (44.6) | 693 (23.1) | 967 (32.3) | 2,996 | 23 (37.7) | 16 (26.2) | 22 (36.1) | 61 |
| Earth Sciences | 446 (40.0) | 282 (25.3) | 386 (34.6) | 1,114 | 21 (42.9) | 10 (20.4) | 18 (36.7) | 49 |
| Biology | 2,027 (40.8) | 1,376 (27.7) | 1,568 (31.5) | 4,971 | 27 (39.1) | 19 (27.5) | 23 (33.3) | 69 |
| Medicine | 4,186 (40.4) | 3,000 (28.9) | 3,177 (30.7) | 10,363 | 22 (32.8) | 21 (31.3) | 24 (35.8) | 67 |
| Agr. & Vet. | 1,235 (40.1) | 660 (21.5) | 1,181 (38.4) | 3,076 | 20 (35.7) | 16 (28.6) | 20 (35.7) | 56 |
| Civil Eng | 604 (39.3) | 326 (21.2) | 605 (39.4) | 1,535 | 22 (40.0) | 13 (23.6) | 20 (36.4) | 55 |
| Ind & Inf Eng | 2,541 (48.9) | 1,211 (23.3) | 1,448 (27.8) | 5,200 | 29 (37.7) | 24 (31.2) | 24 (31.2) | 77 |
| Total | 14,912 (42.8) | 8,951 (25.7) | 10,993 (31.5) | 34,856 | 34 (37.8) | 29 (32.2) | 27 (30.0) | 90 |

*Table 1: Distribution of professors and universities by UDA and geographic macro-region (north, center and south Italy). Percentages are shown in brackets*

## 4. Measuring research performance

We depart from the mainstream and contend that all size-independent indicators based on the ratio to publications, such as the world-renowned Mean Normalized Citation Score, or MNCS, (Waltman et al., 2011) are invalid indicators of performance (Abramo & D'Angelo, 2016a, and 2016b). We measure the research performance of a professor or university by means of a proxy of labor productivity. The calculation of labor productivity requires a few simplifications and assumptions. As a proxy of total output, we consider only the publications (articles, article reviews and proceeding papers) indexed in the Thomson Reuters WoS. When measuring labor productivity, if there are differences in the production factors available to each scientist one should normalize for them. Unfortunately, relevant data are not available at the individual level in Italy. We assume that the same resources are available to all professors within the same field. The second assumption is that the hours devoted to research are more or less the same for all professors. Given the characteristics of the Italian academic system, as noted in Section 2, the above assumptions appear to be acceptable.

---

[3] The complete list is accessible at http://attiministeriali.miur.it/UserFiles/115.htm. Last accessed on September 19, 2016.

Most bibliometricians define productivity as the number of publications in the period of observation. Because publications have different values (impact), we prefer to adopt a more meaningful definition of productivity: the value of output per unit value of labor, all other production factors being equal. Bibliometricians approximate the value or impact with citations. Because citation behavior varies by field, we field-normalize citations. Furthermore, we account for the fractional contributions of scientists to outputs, which is sometimes further signaled by the position of the authors in the list of authors. In Italy, salaries are established at the national level and fixed by academic rank and seniority. Thus, all professors of the same academic rank and seniority receive the same salary, regardless of their merit. Because of the uneven concentration of academic ranks among universities, to avoid distortions in performance measurement at the university level (Abramo et al., 2010), we normalize the individual performance by salary.

At the individual level, we can measure the average yearly productivity, termed fractional scientific strength (*FSS*), as follows[4]:

$$FSS = \frac{1}{w} \cdot \frac{1}{t} \sum_{i=1}^{N} \frac{c_i}{\bar{c}} f_i$$

[1]

Where:
$w$ = average yearly salary of the professor
$t$ = number of years of work of the professor in period under observation
$N$ = number of publications of the professor in period under observation
$c_i$ = citations received by publication $i$
$\bar{c}$ = average of distribution of citations received for all cited publications in same year and subject category of publication $i$[5]
$f_i$ = fractional contribution of the professor to publication $i$.

The fractional contribution equals the inverse of the number of authors in those fields where the practice is to place the authors in simple alphabetical order but assumes different weights in other cases. For the life sciences, widespread practice in Italy is for the authors to indicate the various contributions to the published research by the order of the names in the listing of the authors. For the life science SDSs, we give different weights to each co-author according to their position in the list of authors and the character of the co-authorship (intra-mural or extra-mural) (Abramo et al., 2013a; 2013b).

Because of the differences in the intensity of publications across fields, a prerequisite of any distortion-free performance assessment is to classify each researcher into a single field, or SDS (Abramo et al., 2013c). To compare the research performance of researchers belonging to different fields, or at the aggregate level of universities and territories it is needed to rescale the productivity of individuals by field (SDS), to account for differences of productivity across fields. The standardized productivity of a professor ($FSS^*$) is then defined as:

$$FSS^* = \frac{FSS}{\overline{FSS}}$$

[2]

Where:

$\overline{FSS}$ = national average productivity of all productive researchers in the same SDS of the researcher.

The bibliometric dataset used to measure $FSS$ is extracted from the Italian Observatory of Public Research, a database developed and maintained by the present authors and derived under license from the Thomson Reuters WoS. Beginning from the raw data of the WoS, and applying a complex algorithm to reconcile the author's affiliation and disambiguation of the true identity of the authors, each publication (article, article review and conference proceeding) is attributed to the university scientist or scientists that produced it (D'Angelo et al., 2011). Thanks to this algorithm we can produce ranking lists by $FSS^*$ for each SDS, expressed on a percentile scale of 0-100 (worst to best).

To calculate the yearly research productivity at the aggregate level, be it SDS, UDA, or university, we simply average the yearly standardized productivity of the researchers belonging to the unit. In formulae, the yearly standardized productivity of a university, $FSS^U$ is then:

$$FSS^U = \frac{1}{RS} \sum_{j=1}^{RS} FSS_j^*$$

[3]

Where:

$RS$ = research staff of the university, in the period under observation;

$FSS_{R_j}$ = yearly standardized productivity of researcher $j$ in the university.

Once the research performance has been calculated at the individual and aggregate levels, we will group the professors and universities by macro-region. Previous research has shown that full professors are more productive than associates, and associates more productive than assistants (Abramo et al., 2011b). Therefore, to identify the potential territorial differences of performance in each academic rank, we will also measure the research productivity per academic rank. We will do the same by gender, for similar reasons (Abramo et al., 2015c).

## 5. Results

### 5.1 The research productivity of professors per macro-region

In this subsection, we analyze the 2009-2013 performance of the professors, first in a general analysis, then by SDS and UDA.

The general analysis shows that the average FSS* of professors in the north is 1.01, compared to 0.84 for the professors of the center and 0.75 for the professors of the south (Table 2). The average percentile of FSS* for the professors of the north is 52.0, of those of the center 47.8, and of those of the south 45.6.[6] The share of northern professors that have not published and/or have not received citations for their publications (FSS* = 0) is 10.2% of total, compared to 11.8% of the center professors and 13.3% of the southern professors. In the north, 18.7% of the professors fall within the bottom 20% of national scientists; for the center the share is 22.1%, and for the south it is 24.1%. Some 54.3% of the northern professors have an FSS* above the national median, compared to 47.8% of the center professors and 45.3% of the southern professors. In the north, 23.2% of the professors are within the top 20% of national scientists by FSS*, compared to 19.1% of the center professors and 16.8% of the southern professors. Further, 12.3% of the professors in the north place within the top 10% scientists, while at the center the share is 9.4%, and in the south only 8.0%. In the north, 0.7% of the professors achieve the level of absolute top; for the center the related share is 0.5%, and for the professors of the south the share drops to 0.4%.

|  | North | Center | South | Total |
|---|---|---|---|---|
| Observations | 14,912 | 8,951 | 10,993 | 34,856 |
| % Unproductive professors | 10.2 | 11.8 | 13.3 | 11.6 |
| Average FSS* | 1.01 | 0.84 | 0.75 | 0.88[†] |
| Average percentile rank by FSS* | 52.0 | 47.8 | 45.6 | 48.9 |
| %Bottom 10% | 12.5 | 14.8 | 16.3 | 14.3[††] |
| %Bottom 20% | 18.7 | 22.1 | 24.1 | 21.3 |
| %Above the median | 54.3 | 47.8 | 45.3 | 49.8 |
| %Top 20% | 23.2 | 19.1 | 16.8 | 20.1 |
| %Top 10% | 12.3 | 9.4 | 8.0 | 10.2 |
| %Top | 0.7 | 0.5 | 0.4 | 0.5 |

*Table 2: Descriptive statistics of research performance (FSS*) of the professors of northern, central and southern Italy over the period 2009-2013*

[†] *This value is less than one because for the rescaling of the absolute values of FSS, we use the average without "zero productivity" professors of each SDS, but in calculating the average FSS* we include the unproductive ones.*

[††]*This value is greater than 10% because in each SDS the unproductive professors (FSS* = 0) are more than 10% of the total of scientists in that SDS, and these are all certainly part of the bottom 10%. The same observation applies to the bottom 20%.*

Figure 1 shows the frequency distribution of professors' rank by research productivity in each macro-region. In each class below the median rank, the share of professors in the south is always higher than in the north. The opposite is true above the median. Leaving aside the unproductive professors, the highest frequency for the northern professors is in the (90;100) percentile rank range, while for southern professors it is in the (30;40] range.

As expected, the analysis by gender shows that in general, the average FSS* is higher for the male professors (49.8) than for the female professors (47.0) (Table 3). At the territorial level, the situation by gender is much the same as that seen in the general analysis, with the professors of the north performing more highly than those of the center, who in turn perform higher than those of the south. In the case of the female professors, it seems there is a lesser difference by geographic area, in the rates of the top scientists of northern, central and southern Italy. As for the analysis by gender, the analysis by academic rank shows a situation very similar to what emerges from the general analysis, with the northern professors, independent of their rank, performing above the center professors, who in turn perform above those of the south (Table 4). For the full and associate professors there seems to be a greater difference between the geographic areas in the rate of top scientists, compared to what occurs in the case of the assistant professors, in both cases with greater prevalence of top scientists in the north.

---

[6] Because of its discrete character, the percentile scale may mislead the interpretation of productivity differences between north and south. 25% lower average productivity in the south results in only 6.4 percentile difference.

*Figure 1: Frequency distribution of percentile rank by research productivity (FSS\*) of professors per macro-region*

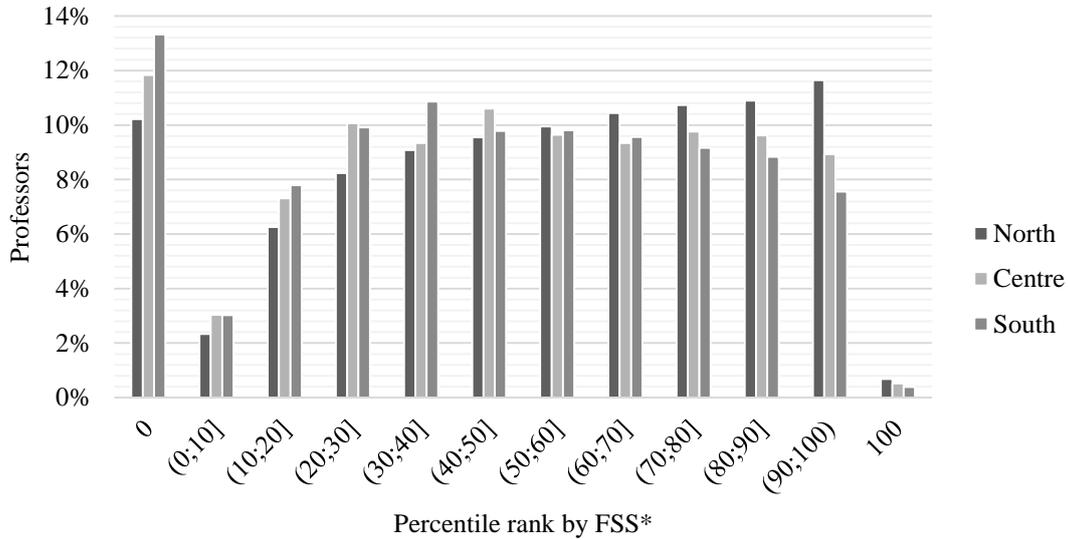

|  | Female | | | | Male | | | |
|---|---|---|---|---|---|---|---|---|
|  | North | Center | South | Total | North | Center | South | Total |
| Observations | 4,562 | 2,844 | 3,382 | 10,788 | 10,350 | 6,107 | 7,611 | 24,068 |
|  | (30.6%) | (31.8%) | (30.8%) | (30.9%) | (69.4%) | (68.2%) | (69.2%) | (69.0%) |
| Aver. percentile by FSS\* | 49.7 | 45.7 | 44.7 | 47.0 | 53.1 | 48.8 | 46.0 | 49.8 |
| % Unproductive professors | 9.2 | 10.7 | 12.1 | 10.5 | 10.7 | 12.4 | 13.9 | 12.1 |
| %Bottom 10% | 11.9 | 14.5 | 15.6 | 13.8 | 12.8 | 15.0 | 16.6 | 14.6 |
| %Bottom 20% | 19.6 | 22.8 | 24.9 | 22.1 | 18.4 | 21.8 | 23.7 | 20.9 |
| %Above the median | 51.6 | 43.9 | 44.7 | 47.4 | 55.5 | 49.6 | 45.6 | 50.9 |
| %Top 20% | 18.5 | 15.6 | 15.1 | 16.7 | 25.3 | 20.6 | 17.5 | 21.7 |
| %Top 10% | 8.4 | 7.4 | 6.7 | 7.6 | 14.0 | 10.4 | 8.5 | 11.4 |
| %Top | 0.3 | 0.3 | 0.2 | 0.3 | 0.9 | 0.6 | 0.5 | 0.7 |

*Table 3: Descriptive statistics of research performance (FSS\*) of the professors of northern, central and southern Italy over the period 2009-2013, by gender*

|  | Assistant professor | | | Associate professor | | | Full professor | | |
|---|---|---|---|---|---|---|---|---|---|
|  | North | Center | South | North | Center | South | North | Center | South |
| Observations | 5,376 | 3,789 | 4,501 | 5,428 | 2,824 | 3,504 | 4,108 | 2,338 | 2,988 |
| Average percentile rank by FSS\* | 49.6 | 45.4 | 44.6 | 53.2 | 49.1 | 45.8 | 53.7 | 50.2 | 47.0 |
| % Unproductive professors | 12.8 | 14.9 | 16.0 | 9.9 | 11.4 | 14.6 | 7.3 | 7.4 | 7.8 |
| %Bottom 10% | 15.3 | 18.0 | 19.1 | 12.3 | 14.5 | 17.3 | 9.3 | 10.1 | 11.0 |
| %Bottom 20% | 21.5 | 25.6 | 27.2 | 18.6 | 21.3 | 24.5 | 15.4 | 17.5 | 18.8 |
| %Above the median | 51.4 | 45.0 | 44.7 | 56.1 | 49.7 | 45.7 | 55.8 | 50.1 | 45.8 |
| %Top 20% | 21.0 | 17.4 | 17.6 | 24.9 | 20.2 | 18.3 | 23.9 | 20.4 | 13.9 |
| %Top 10% | 11.1 | 9.0 | 8.8 | 13.3 | 10.1 | 8.0 | 12.6 | 9.4 | 6.6 |
| %Top | 0.6 | 0.4 | 0.4 | 0.8 | 0.7 | 0.5 | 0.6 | 0.5 | 0.3 |

*Table 4: Descriptive statistics of research performance (FSS\*) of the professors of northern, central and southern Italy over the period 2009-2013, by academic rank*

The analysis by UDA shows that in all the UDAs, the average productivity[7] of the northern professors is greater than that of the southern professors (Table 5).

| UDA | Macro-region | Observations | Average percentile rank by FSS* | % Unproductive prof. | % Bottom 10% | % Bottom 20% | % Above the median | % Top 20% | % Top 10% | % Top |
|---|---|---|---|---|---|---|---|---|---|---|
| Mathematics and computer science | N | 1,522 | 48.4 | 19.1 | 19.3 | 22.5 | 50.9 | 20.5 | 10.9 | 0.3 |
| | C | 807 | 46.5 | 20.8 | 20.9 | 24.9 | 48.1 | 18.3 | 8.9 | 0.2 |
| | S | 939 | 46.5 | 21.7 | 21.8 | 25.0 | 47.8 | 20.8 | 9.7 | 0.4 |
| Physics | N | 1,015 | 51.2 | 7.0 | 10.6 | 19.8 | 52.6 | 22.5 | 11.7 | 0.6 |
| | C | 596 | 49.7 | 5.2 | 9.6 | 19.8 | 48.5 | 19.6 | 10.1 | 0.0 |
| | S | 722 | 47.2 | 8.2 | 11.6 | 22.0 | 47.2 | 17.0 | 7.9 | 0.3 |
| Chemistry | N | 1,336 | 51.2 | 2.2 | 7.8 | 17.7 | 51.3 | 19.9 | 9.8 | 0.3 |
| | C | 693 | 50.0 | 4.2 | 12.6 | 21.6 | 50.6 | 21.5 | 11.5 | 0.4 |
| | S | 967 | 48.1 | 3.7 | 11.6 | 22.3 | 47.3 | 19.3 | 9.5 | 0.4 |
| Earth sciences | N | 446 | 52.2 | 9.0 | 10.8 | 17.0 | 53.4 | 22.0 | 12.6 | 1.1 |
| | C | 282 | 50.0 | 11.7 | 13.5 | 21.3 | 49.3 | 24.1 | 13.5 | 2.1 |
| | S | 386 | 45.5 | 12.2 | 15.3 | 23.6 | 46.1 | 15.3 | 5.4 | 0.3 |
| Biology | N | 2,027 | 52.6 | 5.1 | 9.8 | 18.1 | 54.0 | 23.3 | 12.5 | 0.4 |
| | C | 1,376 | 50.4 | 4.9 | 9.5 | 19.9 | 50.4 | 21.4 | 10.7 | 0.5 |
| | S | 1,568 | 45.6 | 7.0 | 11.9 | 22.6 | 44.2 | 14.6 | 6.5 | 0.2 |
| Medicine | N | 4,186 | 54.5 | 10.1 | 11.4 | 16.4 | 58.1 | 25.8 | 14.3 | 0.8 |
| | C | 3,000 | 47.4 | 12.9 | 15.1 | 21.3 | 47.3 | 18.1 | 8.4 | 0.4 |
| | S | 3,177 | 43.0 | 16.2 | 18.6 | 26.3 | 41.5 | 14.6 | 6.7 | 0.1 |
| Agricultural and veterinary sciences | N | 1,235 | 52.0 | 12.0 | 14.4 | 20.1 | 54.4 | 24.5 | 11.9 | 1.2 |
| | C | 660 | 45.1 | 16.1 | 20.0 | 26.2 | 47.3 | 17.7 | 8.6 | 0.9 |
| | S | 1,181 | 45.7 | 15.0 | 17.9 | 26.7 | 46.1 | 17.3 | 9.7 | 0.8 |
| Civil engineering | N | 604 | 49.9 | 19.9 | 20.2 | 21.4 | 53.5 | 23.0 | 12.3 | 0.7 |
| | C | 326 | 46.2 | 22.1 | 22.1 | 23.3 | 47.2 | 16.6 | 7.1 | 0.0 |
| | S | 605 | 46.0 | 21.8 | 22.0 | 24.3 | 47.8 | 19.0 | 9.9 | 0.8 |
| Industrial and information engineering | N | 2,541 | 50.8 | 11.7 | 13.3 | 20.0 | 53.0 | 22.2 | 11.5 | 0.8 |
| | C | 1,211 | 46.0 | 13.6 | 15.7 | 23.8 | 44.1 | 17.8 | 9.7 | 0.9 |
| | S | 1,448 | 48.3 | 12.8 | 14.7 | 20.3 | 49.0 | 18.6 | 8.6 | 0.8 |

*Table 5: Descriptive statistics of research performance (FSS\*) of the professors of northern (N), central (C) and southern (S) Italy over the period 2009-2013, by UDA*

As well, there are only two UDAs (Industrial and information engineering; Agricultural and veterinary sciences), where the southern professors seem on average to be higher performing than those of the center. The greatest difference in average productivity by FSS* between northern and southern professors occurs in Medicine (54.5 average productivity of the northern professors, against 43.0 for those in the south). The smallest differences occur in Mathematics and computer science (48.4 for the northern professors against 46.5 for those of the south) and Industrial and information engineering (50.8 in the north against 48.3 in the south). Further, the rate of unproductive professors is always greater in the south than in the north. Still, in two UDAs (Physics; Biology), the rate of unproductive professors in the north is greater than that of the center; and in a full four UDAs (Agricultural and veterinary sciences; Industrial and information engineering; Chemistry; Civil engineering), the rate of unproductive professors in the center is greater than that of the south. Concerning the ratio of the scientists in the national top 20% to the total of scientists for the macro-region, this is always higher in the north, with the exception of the Mathematics and computer science (where the higher ratio occurs in the south) and Chemistry and earth sciences UDAs (where we find the highest ratio in central Italy).

---

[7] From this point, for "average productivity" we will use the average percentile rank by FSS rather than the average FSS, which would be affected by the presence of outliers.

In Table 6, we present the descriptive statistics for the research productivity differentials, comparing the professors of the north, center and south at the SDS level. The gaps of average percentile by FSS* between the professors of the three geographic areas are presented by paired comparisons: north vs south, north vs center, center vs south.

The analysis excludes all the SDSs (10 in total) that have less than three observations in any of the three geographic areas.[8] Following these exclusions, the dataset for the analysis consists of 182 different SDSs.

In the case of the north-south comparison, the number of SDSs with a positive gap is 140 out of 182 (76.9%). The highest positive gap is for ING-IND/34 (Industrial bio-engineering) (35.0 percentiles of FSS*); the highest negative gap is for AGR/06 (Wood and forest uses technology) (-47.0 percentiles of FSS*).

In the case of the center-south comparison, the number of SDSs with positive gap is 105 out of 182 (57.7%). The highest positive gap is for ING-IND/29 (Raw materials engineering) (47.9 percentiles of FSS*); the highest negative is for ING-IND/28 (Excavations engineering and safety) (-33.3 percentiles of FSS*).

|  | North - South gap | North - Center gap | Center - South gap |
|---|---|---|---|
| Highest positive gap | 35.0 (ING-IND/34) | 39.3 (ING-IND/28) | 47.9 (ING-IND/29) |
| Highest negative gap | -47.0 (AGR/06) | -42.5 (ING-IND/29) | -33.3 (ING-IND/28) |
| Number of SDSs with gap >=0 | 140 (76.9%) | 134 (73.6%) | 105 (57.7%) |
| Number of SDSs with gap <0 | 42 (23.1%) | 48 (26.4%) | 77 (42.3%) |

*Table 6: Descriptive statistics of research performance (FSS*) of the professors of northern, central and southern Italy over the period 2009-2013, by SDS*

## 5.2 The research productivity of universities per macro-region

In this sub-section, we analyze the research performance of the universities grouped by macro-region, both at the general level and at the UDA and SDS levels.

The general analysis (Table 7, last row) shows that the average percentile of $FSS^U$ of the 26 northern universities is 61.1, compared to 53.2 for the 16 central universities and 34.6 for the 22 southern universities. The share of northern universities in the bottom 20% for the nation is 7.7%, for the central universities the share is 12.5%, and for the southern universities it is a full 40.9%. Only 31.8% of the southern universities have performance above the median, compared to 65.4% for the northern universities and 50.0% of those in the center. Further, only 4.5% of the southern universities place among the top 20% of universities, against 30.8% of the northern universities and

---

[8] The SDSs excluded are: CHIM/05, FIS/08, GEO/12, ING-IND/18, ING-IND/20, ING-IND/30, MED/47, ING-IND/01, ING-IND/02, ING-IND/23.

25.0% of those in the center. Finally, no southern university places among the top 10% of universities, against 18.8% of the central universities and 15.4% of those of the north. As we might have expected, this first analysis reveals that when we shift the focus from the professors to the universities, the difference by geographic area seems much more relevant, with the universities of the south performing much lower at the national level than did the individual professors of the south.

| UDA | Macro-region | Observations* | Average percentile rank by FSS$^U$ | %Bottom 10% | %Bottom 20% | %Above the median | %Top 20% | %Top 10% | %Top |
|---|---|---|---|---|---|---|---|---|---|
| Mathematics and computer science | N | 24 | 52.3 | 8.3 | 16.7 | 58.3 | 16.7 | 4.2 | 0.0 |
| | C | 10 | 42.4 | 10.0 | 10.0 | 30.0 | 20.0 | 10.0 | 0.0 |
| | S | 16 | 51.3 | 12.5 | 31.3 | 50.0 | 25.0 | 18.8 | 6.3 |
| Physics | N | 20 | 53.5 | 15.0 | 20.0 | 50.0 | 20.0 | 15.0 | 0.0 |
| | C | 10 | 51.4 | 10.0 | 20.0 | 50.0 | 30.0 | 10.0 | 10.0 |
| | S | 13 | 43.6 | 7.7 | 23.1 | 46.2 | 15.4 | 7.7 | 0.0 |
| Chemistry | N | 17 | 55.2 | 11.8 | 11.8 | 58.8 | 23.5 | 11.8 | 0.0 |
| | C | 10 | 51.2 | 20.0 | 20.0 | 50.0 | 20.0 | 20.0 | 0.0 |
| | S | 15 | 43.3 | 6.7 | 33.3 | 40.0 | 20.0 | 6.7 | 6.7 |
| Earth sciences | N | 11 | 58.2 | 0.0 | 9.1 | 54.5 | 36.4 | 18.2 | 0.0 |
| | C | 8 | 59.2 | 0.0 | 0.0 | 62.5 | 25.0 | 12.5 | 12.5 |
| | S | 12 | 36.4 | 25.0 | 41.7 | 33.3 | 0.0 | 0.0 | 0.0 |
| Biology | N | 20 | 65.4 | 5.0 | 5.0 | 75.0 | 25.0 | 15.0 | 5.0 |
| | C | 14 | 57.3 | 0.0 | 7.1 | 57.1 | 35.7 | 21.4 | 0.0 |
| | S | 19 | 28.4 | 26.3 | 47.4 | 15.8 | 5.3 | 0.0 | 0.0 |
| Medicine | N | 18 | 68.0 | 0.0 | 5.6 | 77.8 | 38.9 | 27.8 | 5.6 |
| | C | 9 | 49.9 | 11.1 | 11.1 | 55.6 | 22.2 | 0.0 | 0.0 |
| | S | 15 | 28.5 | 26.7 | 46.7 | 13.3 | 0.0 | 0.0 | 0.0 |
| Agricultural and veterinary sciences | N | 10 | 57.1 | 0.0 | 20.0 | 70.0 | 20.0 | 10.0 | 10.0 |
| | C | 8 | 48.2 | 12.5 | 12.5 | 25.0 | 25.0 | 12.5 | 0.0 |
| | S | 11 | 44.8 | 18.2 | 27.3 | 45.5 | 18.2 | 9.1 | 0.0 |
| Civil engineering | N | 14 | 59.4 | 7.1 | 14.3 | 64.3 | 21.4 | 14.3 | 0.0 |
| | C | 8 | 51.4 | 0.0 | 12.5 | 62.5 | 25.0 | 12.5 | 0.0 |
| | S | 14 | 39.8 | 21.4 | 28.6 | 28.6 | 14.3 | 7.1 | 7.1 |
| Industrial and information engineering | N | 19 | 43.5 | 21.1 | 26.3 | 42.1 | 15.8 | 5.3 | 0.0 |
| | C | 12 | 52.4 | 8.3 | 25.0 | 50.0 | 25.0 | 16.7 | 0.0 |
| | S | 18 | 55.2 | 0.0 | 11.1 | 55.6 | 22.2 | 11.1 | 5.6 |
| Overall** | N | 26 | 61.1 | 0.0 | 7.7 | 65.4 | 30.8 | 15.4 | 3.8 |
| | C | 16 | 53.2 | 6.3 | 12.5 | 50.0 | 25.0 | 18.8 | 0.0 |
| | S | 22 | 34.6 | 27.3 | 40.9 | 31.8 | 4.5 | 0.0 | 0.0 |

*Table 7: Macro-regional analysis of the research performance (FSS$^U$) of the universities at the UDA level*
* The analysis by UDA excludes the universities with less than five professors in the UDA.
** In the overall analysis, the field of observation excludes the SDS-university pairs with less than 10 professors in the SDS.

The analysis by UDA shows that in all the UDAs, the average percentile of FSS$^U$ of the northern universities is greater than of the southern universities, except for Industrial and information engineering, where the performance of the southern universities (55.2) is superior to that of the both the central universities (52.4) and the northern universities (43.5). Apart from Industrial and information engineering, the southern universities also register a higher performance than those of the center in Mathematics and computer science (51.3 against 42.4). The greatest difference in average productivity between north and south is in Medicine (the average percentile for the northern universities is 68.0, against 28.5 for those of the south) and in Biology (65.4 against 28.4). Concerning the share of the universities of a macro-region in the bottom 20% for the nation, this is always greater in the south than in both in the north and center, with the lone exception of Industrial and information engineering.

Concerning the share of universities reaching the top 20% for the nation, there are five UDAs where the central universities have the largest share (Biology; Physics; Civil engineering; Industrial and information engineering; Agricultural and veterinary sciences), three UDAs where the northern universities achieve the largest share (Medicine; Earth sciences; Chemistry); and only one UDA where the southern universities have the largest share (Mathematics and computer science).

Table 8 presents the descriptive statistics of the distributions of $FSS^U$ differences between the universities of northern, central and southern Italy, at the level of SDS.

The analysis excludes all those SDSs that have less than three universities per geographic area, as well as those individual university SDSs with less than three research staff. Following these exclusions, the dataset for the analysis is composed of 132 different SDSs.

In the case of the north-south comparison, the number of SDSs with positive gap is 97 out of 132 (73.5%). The highest positive gap is in MAT/06 (Mathematical probability and statistics) (47.4 percentiles of $FSS^U$); the highest negative gap is in VET/06 (Parasitology and parasitic diseases of animals) (-36.4 percentiles of $FSS^U$).

In the case of the center-south comparison, the number of SDSs with positive gap is 85 out of 132 (64.4%). The highest positive gap is again in MAT/06 (64.3 percentiles of $FSS^U$), and the highest negative gap is again in VET/06 (-38.6 percentiles of $FSS^U$).

|  | North-South gap | North-Center gap | Center-South gap |
|---|---|---|---|
| Highest positive gap | 47.4 (MAT/06) | 50.1 (ING-IND/14) | 64.3 (MAT/06) |
| Highest negative gap | -36.4 (VET/06) | -31.8 (MED/39) | -38.6 (VET/06) |
| Number SDS with gap >=0 | 97 (73.5) | 86 (65.2) | 85 (64.4) |
| Number SDS with gap <0 | 35 (26.5) | 46 (34.8) | 47 (35.6) |

*Table 8: Descriptive statistics of the research productivity differences $FSS^U$, at the SDS level, of the universities of northern, central and southern Italy over the period 2009-2013*

## 6. Discussion

In this section, we propose some potential explanations of the results, with the objective of providing light on the possible causes of the north-south differential in the Italian higher education system. Some of these hypotheses will be supported by the data and by other studies already presented in the literature. Others will for the moment remain hypotheses, given the current impossibility of testing them and the scarcity of other supporting studies.

Our results show the unequivocal existence of a differential in research performance in the sciences between the professors of northern, central and southern Italy. In particular, the professors of the north result as on average higher performing than those of the center, who are in turn higher performing than those of the south. Analogous differences in performance also occur in the analysis of the share of unproductive professors (higher in the south), and of top scientists (higher in the north). This trend does not seem to vary notably when we deepen the analysis by gender and academic rank. The analysis by gender shows that in the case of the female professors, there seems to be a lesser difference by geographic area in the share of the top scientists. The analysis by academic rank shows that in the case of the full and associate professors, the difference by geographic area is higher than for the assistant professors, as concerns the share of top scientists. Since the assistant professors represent the new generation of academics, this finding could offer some comfort for the future. The analysis at the aggregate level of the university shows a north-south difference that is even greater than that for the professors.

The origins of the north-south gap could have different roots, of economic, social, cultural and historical-geographical character. As observed by Sánchez-Barrioluengo (2014), "not only does the university influence the surrounding region, but also regional characteristics shape university performance".

At the macroeconomic level, Aiello and Scoppa (2006) show that in general, a large part of productivity gaps across Italian regions cannot be imputed to differences in physical or human capital, but rather are related to relevant differences in total factor productivity (TFP), which depends on how efficiently and intensely the inputs are utilized in production. In a previous study, Aiello and Scoppa (2000) had indicated the following variables as having important influence on TFP, and so on the Italian regional differences: infrastructure, state intervention, the financial system, and property rights enforcement. However, in a subsequent work, Scoppa (2007) re-evaluates the importance of human and physical capital, rather than TFP, as an explanation of the uneven development across Italian regions in the period 2000-2004. In this study, he finds that both human and physical capital are unequally distributed and highly correlated with regional productivity.

The greater industrialization and economic development of northern Italy could thus have contributed to the scores of research performance which we have measured. According to this hypothesis, over time, the greater economic development of the north could have favored the formation of collaborations between researchers and local industry, stimulating the originality, applicability and innovation of the research, and generating private financing and support for academic activity. As shown by Abramo et al. (2011b), for Italian industry, geographic proximity plays a relevant role in Italy in the choice of their academic partners in any joint research projects. Research collaboration with industry then seems to have a positive impact on academic research performance (Balconi and Liberanti, 2006). Furthermore, the greater economic development of northern Italy may have not only brought greater private financing to research activity, but also greater public financing. In fact, as well as the GDP of the north being higher than for the south, R&D expenditures as a percentage of GDP are also higher in the north than in the south (ISTAT, 2015b). More specifically, since 2009, a share of government financing has been distributed to universities in function of the outcomes of the national research evaluation exercises (VTR 2001-2003, VQR 2004-2010). Since the northern universities on average resulted as higher performing than those of the south, the latter were penalized by the PBRF adopted by the government. As a matter of fact, between 2008 and 2014, public funds allocated by the governments to universities (FFO), diminished by 0.7% in the north and by 10.7% in the south. Average tuition fees in the south are about 50% of those in the north (ANVUR 2014, p. 213 and Viesti, 2015). Analogous are regional disparities in R&D expenditures (ANVUR, 2016, p. 468). In the period 2009-2012, the receipts of universities per professor in the south were 79% of those in the north. The receipts per student in the south were 67% of those of in the north (ANVUR 2014; Banfi & Viesti, 2015). Our proxy indicator of labor productivity, *FSS*, assumes that resources to individuals are the same, therefore it does not control for unequal production factors. A part of the performance differential registered could thus be caused by the greater resources available and the opportunities for collaboration with industry in the north.

The north-south economic gap widened further in connection with the financial and economic crisis that played out in Italy over the years 2008 to 2014, and which hit the southern regions much more than those in the north. Tuition fees in southern universities increased more than the national average; the family income decreased more than the national average; and the share of employed graduates decreased more than elsewhere. Many families in southern Italy saw their total income fall below the minimum level necessary to enroll their children in university. This would have caused fewer students in the ranks of the southern universities, worsened still further by the net migration of the more well-off students towards the northern and central universities (ANVUR, 2014, p. 37). Nevertheless, professors in the south still bear a slightly higher teaching load than those in the north (ANVUR, 2016, p. 406). A four-hour teaching load difference per academic year, though, can hardly justify the conspicuous research productivity gap. Whereas, less income from tuition fees and per-student state funding often translates into less allocation for academic research.

The migratory phenomenon is also taking place among the best professors (Francalacci, 2015; Cappelletti Montano, 2015), who are attracted by the better universities, the more stimulating environment, and the generally better quality of life offered by the northern regions (ISTAT, 2015c). This brain drain of students and professors from south to north will likely provoke further cultural impoverishment of the south, creating a vicious circle leading to the exodus of further competencies and capacities from south to north, and a still greater north-south differential.

Among the possible socio-cultural causes of this gap, there is a frequently encountered opinion that favoritism in academic institutions is more concentrated in southern than northern Italy (Allesina 2011; Durante et al. 2011). Abramo et al. (2015a), studying the determinants of academic career advancement in Italy, found a weak positive association between the "expected" outcomes of competitions (i.e. the winner truly has scientific merit) and the fact that these are held in northern universities rather than those in southern or central Italy. Abramo et al. (2014c) show that the distribution of ''parent-child'' professor pairs by geographic area of university location shows the maximum concentration in southern Italy, while the minimum concentration is in the north, and identify one of the causes of this distribution as possibly greater diffusion of the phenomenon of nepotism in the south. However, these same authors do not find statistically significant differences in performance between "children" and "non-children" professors by geographic area, with the exception of central Italy, where the children on average result as more productive than the non-children. Although possible favoritism among the non-children sub-population was not investigated, in general, the evidence emerging from the study of phenomena of nepotism and favoritism does not seem sufficient to explain the north-south gap. From a general perspective, it is also possible to compare the university effectiveness in recruitment, turnover and overall mobility, at the geographic level. The current authors have conducted this precise inquiry for the period 2008 to 2012, on the basis of the methodology described in Abramo et al. (2016). The results obtained are contrasting, in the sense that do not permit the establishment of a direct causal relationship with the north-south gap that emerges in the current study. Above all, nothing can be concluded about the effectiveness of recruitment and turnover in the preceding years, which certainly could have had some influence on the performance of the current period, now under examination.

A further cause of the north-south gap could be linked to the different gender concentration among the professors. In fact, the scientific literature shows a gap in performance between the researchers of male and female gender (Abramo et al., 2015b; Abramo and D'Angelo, 2015b; Larivière et al., 2013; Mauleón and Bordons, 2006; Xie and Shauman, 2004; Long, 1992; Fox, 1983). Still, analyzing our dataset, this hypothesis does not receive empirical confirmation. Indeed, in general, the incidence of women in the total research staff does not differ much between the north (where women are 30.6% of the research staff) and the south (30.8%). For the Medicine UDA, which is the one with the greatest detected difference in average percentile between the northern professors and universities and those of the south, the opposite seems true: in the northern universities, the incidence of women in the university research staff is higher than it is in the south.

A further cause of the performance differential could derive from structural factors that influence the university productivity. Observing the average and median size of the university SDSs, we note that it is higher in the northern universities (average research staff per SDS = 6.61) compared to the southern universities (5.51). This occurs in every UDA, with the maximum difference in Chemistry (10.80 vs 7.76) and the minimum in Civil engineering (4.78 vs 4.50).

However, various studies have excluded positive returns to size in research (Seglen and Aksnes, 2000; Golden and Carstensen, 1992), in particular in Italy (Abramo et al., 2012c; Bonaccorsi and Daraio, 2005), as well as excluding returns to scope (Abramo et al., 2014d). On the basis of the results emerging in the literature, we can therefore exclude that the size of the research staff could have any impact on the research performance gap between northern and southern Italy.

Other factors that could influence the north-south gap might be linked to other cultural, sociological, or even climatic aspects, which are mentioned in the literature. However, we are not able to empirically verify the incidence of these factors.

Our own opinion is that various causes (including those that we have been able to empirically verify in the current paper) could have interacted and contributed to one another, and through a cumulative process, brought about the gap detected in this study. As a result, if our hypothesis is correct, the reduction of the north-south gap could not be achieved by concentrating on a few determinative issues, but would require simultaneous action on a number of factors, installing a gradual process that would take into account the multiple socio-economic-cultural variables that characterize the north-south differential.

## 7. Conclusions

The north-south gap in research productivity of Italian professors and universities in the sciences is a fact. The causes could be of an economic nature (less resources available for researchers in the south, compared to those in the north), but this would not be a sufficient explanation. The questions that need response are: i) given this divide, is the policy of university resource allocation on the basis of merit, initiated by the government in 2009, appropriate? ii) what initiatives can be undertaken to reduce this gap, while avoiding prejudice to the improvement of the entire higher education system?

It is widely understood that in the knowledge-based economy, a strong higher education system provides a decisive contribution to the competitiveness and growth of the national economic system, as well as to social mobility. On the one hand, PBRF policies can certainly be functional in reinforcing Italian higher education, serving as a stimulus to improvement. On the other hand, they could contribute to increasing the existing north-south gap, both in higher education and at the macroeconomic and social levels. Moreover, PBRF could undermine the basic social principle of the Italian academic system, spelled out in the Italian Constitution: all students must be guaranteed opportunity of access to equal quality university education, independent of personal standing or geographic location.

The challenge for the policy maker is how to introduce competitive mechanisms and incentive systems that could stimulate the upgrading of the country's universities, while at the same time reducing the qualitative gap between the northern universities and those of the south.

In Italy, an increasing share of total funds is allocated to universities on the basis of the results of periodic national research assessment exercises. Such exercises do not measure individual performance, therefore the allocation is based on the average performance of universities. However, in Italy the variability of research performance is much higher within universities than between universities (Abramo et al., 2012a), and it could therefore happen that top performers in low-tier universities (mainly in the south) would receive fewer funds than low performers in high-tier universities (mainly in the north).

The paradox could be solved providing individual evaluations and having universities allocate resources internally on the basis of the individual performance. Alternatively, research funds could be directly allocated to individual researchers according to their merit. A further step would be to link the professors' salaries to the quality of their teaching and research, as is the case in competitive higher education systems. Liberation of significant resources for universities in the south, given the performance distribution of their professors, could be achieved by cutting faculty with low productivity. This would entail changing the overall context, which is strongly structured against any form of forced departure, and where suspension for low productivity is still almost inconceivable. A final step would be designing policies to foster the concentration of high performers in few universities in the south, through different forms of incentives, including salary differentiation.

Given the fundamental socio-economic role of the universities within their territories, reducing the north-south gap in higher education (and the overall education system) is a necessary structural intervention, for the hope of also reducing the national socio-economic divide.